\documentclass[reprint,amsmath,smsymb,aps,superscriptaddress,aip,apl,nofootinbib]{revtex4-2}
\usepackage{geometry}
\renewcommand{\thesubsection}{\alph{subsection}}
\usepackage{titlesec}
\usepackage{bm}
\usepackage{gensymb}
 \usepackage{graphicx}
 \usepackage{tensor}
 \usepackage{verbatim}
 \usepackage{courier}
 \usepackage{mathtools}
  \usepackage[inline]{enumitem}
 \usepackage{calc}
\graphicspath{}
\usepackage{color}
\usepackage{amsmath}
\usepackage{xfrac}
\usepackage[table]{xcolor}

\begin{document}
\title{When will we observe binary black holes precessing?}
\author{Stephen Fairhurst}
\affiliation{School of Physics and Astronomy, Cardiff University, Cardiff, CF24 3AA, United Kingdom}
\author{Rhys Green}
\affiliation{School of Physics and Astronomy, Cardiff University, Cardiff, CF24 3AA, United Kingdom}
\author{Mark Hannam}
\affiliation{School of Physics and Astronomy, Cardiff University, Cardiff, CF24 3AA, United Kingdom}
\author{Charlie Hoy}
\affiliation{School of Physics and Astronomy, Cardiff University, Cardiff, CF24 3AA, United Kingdom}
\date{\today}

\begin{abstract}
After eleven gravitational-wave detections from compact-binary mergers, we are yet to observe the striking
general-relativistic phenomenon of orbital precession. Measurements of precession would provide valuable
insights into the distribution of black-hole spins, and therefore into astrophysical binary formation mechanisms.
Using our recent two-harmonic approximation of precessing-binary signals~\cite{Fairhurst:2019_2harm}, we introduce the
``precession signal-to-noise ratio'', $\rho_p$. We demonstrate that this can be used to clearly identify whether precession
was measured in an observation (by comparison with both current detections and simulated signals), and
can immediately quantify the measurability of precession in a given signal, which currently requires computationally
expensive parameter-estimation studies. $\rho_p$ has numerous potential applications to signal searches,
source-property measurements, and population studies. We give one example: assuming one possible astrophysical
spin distribution, we predict that precession has a one in $\sim 25$ chance of being observed in
any detection.
\end{abstract}

\maketitle
The Advanced Laser Interferometer Gravitational-Wave Observatory (aLIGO)~\cite{TheLIGOScientific:2014jea} and
Advanced Virgo (AdV)~\cite{TheVirgo:2014hva}, provide a unique method of observing mergers of black holes and/or neutron
stars. Observations to date already provide insights into the mass and spin distributions of black
holes~\cite{LIGOScientific:2018mvr, LIGOScientific:2018jsj}. 

One important general relativistic effect that has
\emph{not yet} been observed is orbital precession. This arises when the black-hole spins are not aligned with the
binary's orbital angular momentum. In contrast to Newtonian mechanics, where all angular momenta are individually
conserved, in general relativity the binary's total angular momentum is (approximately) conserved, and the orbital angular
momentum (and hence the orbital plane) and spins precess around it~\cite{Apostolatos:1994mx, Apostolatos:1995pj}.
This leads to modulations in the amplitude and phase of the gravitational wave (GW) signal. These are in general small
effects and, in addition, whether they can be measured depends not only on the black-hole masses and spin magnitudes and 
directions, but also on the binary's orientation relative to the detector, and the observed GW polarization. For this 
reason, until now there was no straightforward way to determine how significantly precession would be imprinted onto 
a given waveform. The usual approach is to perform computationally expensive Bayesian analyses 
(see e.g. \cite{LIGOScientific:2018mvr, Veitch:2014wba}), but even then,
the misaligned spin components (which signify whether the binary is precessing) are degenerate with other parameters, 
and do not provide a direct measure of precession features in the signal. This makes it difficult to infer the impact of precession measurements on the properties of astrophysical binary populations and their formation mechanisms. 

\textbf{The two-harmonic approximation:}
We take advantage of a new characterization of precession, where the gravitational waveform for a precessing binary can 
be expressed as the sum of five harmonics, which form a power series in%
\begin{equation}
	b = \tan(\beta/2)
\end{equation}
where $\beta$ is the opening angle between total and orbital angular momentum~\cite{Fairhurst:2019_2harm}.  Large values
of $\beta$ are possible only when the in-plane component of spin, $\chi_{p}$~\cite{Schmidt:2014iyl} is large; the mass ratio 
of the system is large, or
there is a significant spin component anti-aligned with the orbital angular momentum --- both of which serve to reduce the
relative magnitude of the orbit-aligned angular momentum.  For most
configurations $b \ll 1$, and we may restrict attention to the first two harmonics and ignore terms of order $b^{2}$ or higher.
It is only for binaries with $\beta \ge 45\degree$, or equivalently $b \ge 0.4$, where the remaining harmonics become 
significant, 
and the two-harmonic approximation breaks down.

We may now write the observed waveform as a function of GW frequency as,
\begin{equation}\label{eq:h_2harm}
	h_{\mathrm{2 harm}}(f) \approx \mathcal{A}_{0} h^{0}(f) + \mathcal{A}_{1} h^{1}(f) \, ,
\end{equation}
where $\mathcal{A}_{0}$ and $\mathcal{A}_{1}$ are complex, orientation dependent amplitudes and $h^{0}(f)$ and $h^{1}(f)$
are the waveforms of the two harmonics. Full details of the two-harmonic approximation are given in 
Ref.~\cite{Fairhurst:2019_2harm}. What is relevant to the current work is that the individual harmonics do not exhibit precession
modulations; these show up due to beating between the harmonics, whose respective dephasing can be related to 
the frequency with which the orbital angular momentum precesses around the total angular momentum. 
The amplitudes $\mathcal{A}_{0}$ and $\mathcal{A}_{1}$
depend on the binary's orientation and polarisation, but are approximately constant throughout a given signal. 

\textbf{Observability of precession:}
Since the individual harmonics are indistinguishable from non-precessing waveforms, it
is only when two precession harmonics can be independently observed that precession can be unambiguously 
identified.  For precession to be observable, we therefore 
require that the expected signal-to-noise ratio (SNR) in \emph{both} of the harmonics is above some threshold.

The expected SNR for a signal $h$ embedded in data from a detector with a noise
power spectral density $S(f)$ is given as $\hat{\rho}^{2} = (h | h)$, where the inner product is,

\begin{equation}
	(g | h) = 4 \, \mathrm{Re} \int_{f_{o}}^{f_{\mathrm{max}}} \frac{g(f) h^{\star}(f)}{S(f)} df \, ,
\end{equation}
and ${}^{\star}$ denotes complex conjugation \cite{Allen:2005fk}.  
In cases with more than one precession cycle between $f_{o}$ and $f_{max}$, 
the two harmonics will be close to orthogonal, $(h^{0} | h^{1}) \approx 0$,
and the precession SNR is simply defined as
the expected SNR in the weaker harmonic:%
\begin{equation}\label{eq:rho_p}
	\rho_{\mathrm{p}} = \mathrm{Min} ( |\mathcal{A}_{0} h^{0} |,  |\mathcal{A}_{1} h^{1} | ) \, ,
\end{equation}
where $|h| := \sqrt{(h | h)}$.  When there is less than one precession cycle, it is necessary to also subtract
the power aligned with the dominant harmonic, as described in detail in Ref.~\cite{Fairhurst:2019_2harm}, before
determining whether the second harmonic is observable.

\begin{figure*}[t]
	\centering
	\includegraphics[scale=0.265]{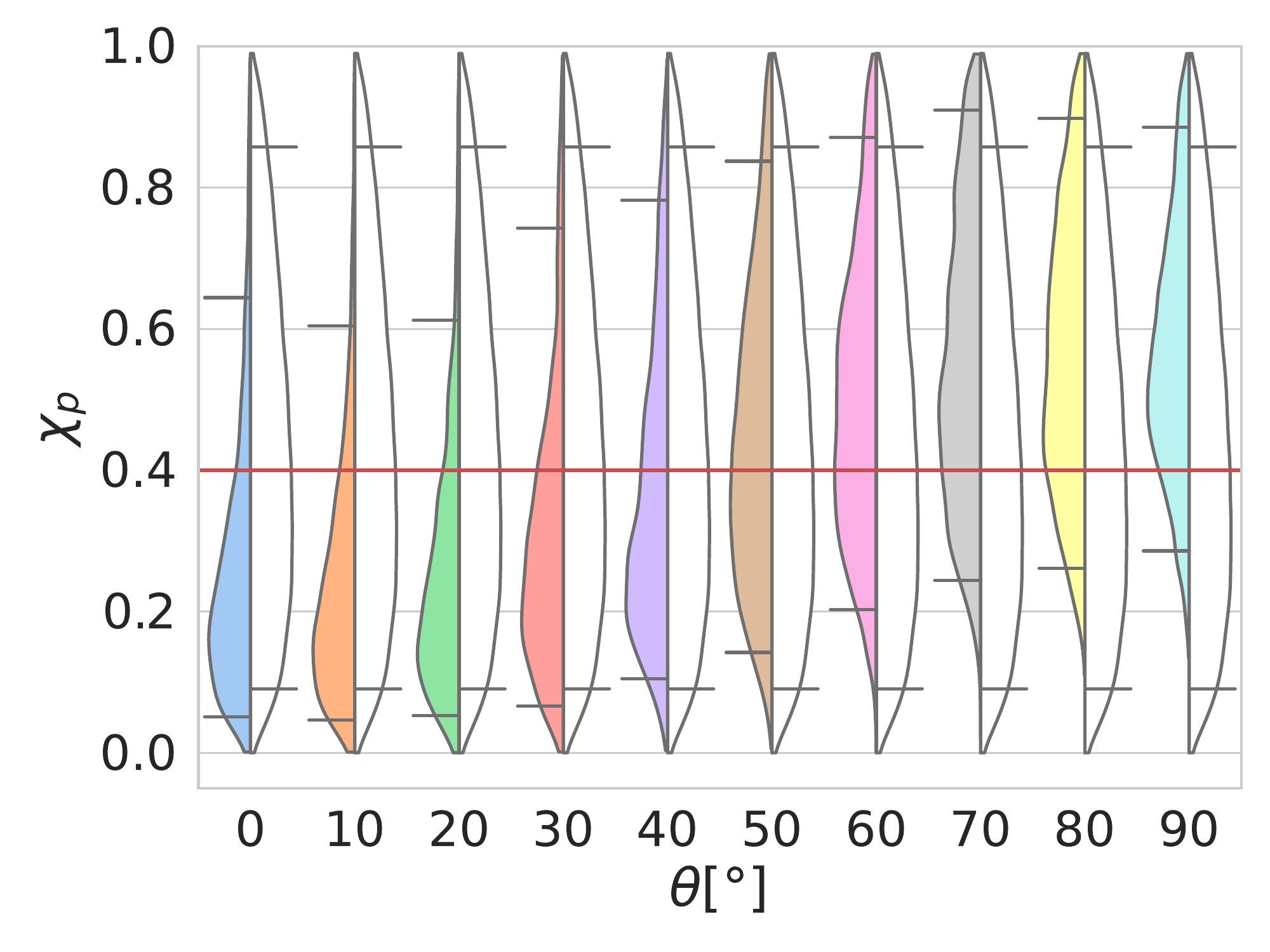}
	\includegraphics[scale=0.26]{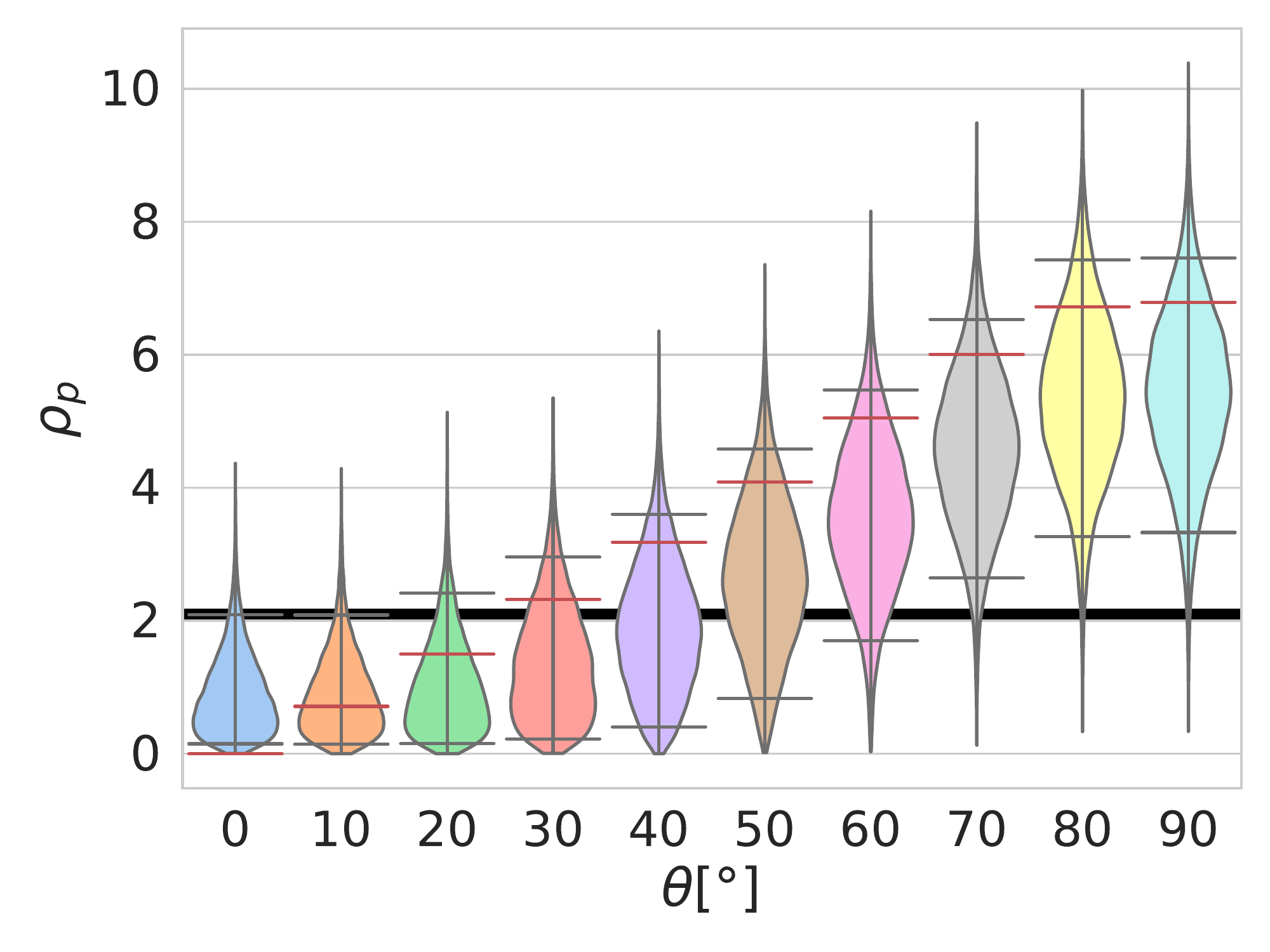}
	\includegraphics[width = 0.65\columnwidth , height = 4cm]{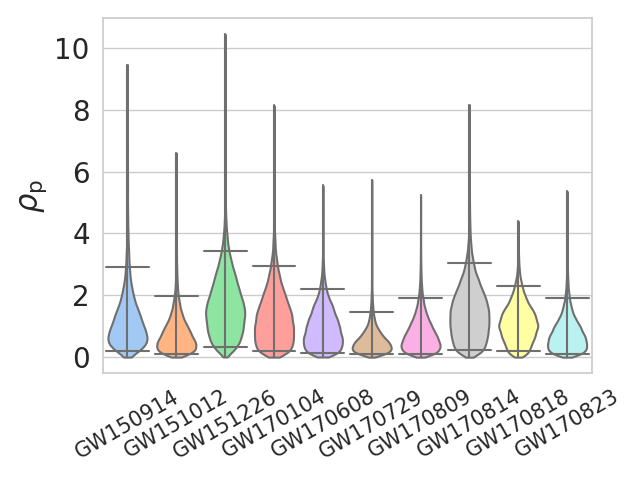}
	\caption{
	For a set of simulated signals with fixed masses and spins (see text), we show the posterior and prior (white) 
	distributions for $\chi_p$ (left), and posterior distributions for $\rho_p$ (middle) for a range of different binary orientations, 
	$\theta$. The grey lines show the $90\%$ confidence regions, the solid red lines show the \textit{true} values of $\chi_{p}$ 
	and $\rho_{p}$ respectively and the thick black line indicates the threshold for observable precession i.e., $\rho_{p} > 2.1$.
	The right panel shows the $\rho_{p}$ distribution for the ten binary-black-hole observations in O1 and 
	O2~\cite{LIGOScientific:2018mvr}.
	}
	\label{fig:chi_p_and_rho_p}
\end{figure*}

It remains to determine a threshold above which $\rho_{p}$ can be considered as evidence for precession.  Consider the situation
where an event has been observed, so there is significant SNR in at least one harmonic.  In the absence of measurable 
precession, and assuming well-modelled Gaussian noise, the SNR in the second harmonic will be $\chi^{2}$ distributed
with two degrees of freedom, where the two degrees of freedom correspond to the real and imaginary parts of the complex
amplitude.  Therefore, in the absence of precession, $\rho_{p}>2.1$ is expected in less than 10\% of
cases, and we use this as a simple threshold for observable precession.%
\footnote{A more detailed analysis would consider the volume of the binary
parameter space 
consistent with the non-precessing and precessing parameters and use these
to appropriately weight the likelihoods of the precessing and non-precessing signals.  
This will have some impact on the required $\rho_{p}$ threshold, but is unlikely to change it significantly.}

In Fig. ~\ref{fig:chi_p_and_rho_p} we show the recovered distribution of $\chi_{p}$ and $\rho_{p}$ for
a number of signals, both real and simulated.  For each signal, we use a nested sampling
routine within the LALInference code \cite{Veitch:2014wba} to
obtain posterior probability distributions for the parameters. This is the same infrastructure that was used to measure the 
properties of the LIGO-Virgo observations, and we present our results in the same form as in, for example, the GWTC-1
catalogue~\cite{LIGOScientific:2018mvr}, by using the PESummary library \cite{Hoy:2020vys}. The new feature is our calculation of $\rho_p$. 

First, we show the recovered $\chi_{p}$ and $\rho_{p}$ 
distributions for a set of simulated signals, generated using the {\tt IMRPhenomPv2} model~\cite{Hannam:2013oca}, each
with the same choices of masses and spins  --- total mass $M=40 M_{\odot}$, mass ratio 2:1, and an in-plane spin of
$\chi_{p} = 0.4$ on the large black hole only --- but varying orientation, encoded by the angle $\theta$ between the total 
angular momentum and the line of sight . The distance to each
signal is chosen to ensure a \textit{fixed} expected SNR of 20 in the aLIGO detectors at the sensitivity
of the second observing run (O2)~\cite{LIGOScientific:2018mvr}, resulting in a distance variation by a factor of
$\approx 3.5$ between the least and most inclined systems.

For binaries with total angular momentum closely aligned with the line of sight, $\theta < 45\degree$,
 the precessing SNR is consistent with no power in the $h^1$.  The posterior on 
$\chi_{p}$ is consistent with the prior at low $\chi_{p}$ but excludes $\chi_{p} \gtrsim 0.7$. When $\theta > 45\degree$, the angular 
momentum is significantly mis-aligned with the line of sight and there is significant power 
in both harmonics, leading to a value of $\rho_{p}$ inconsistent with noise alone and little support for values of $\chi_p \lesssim 0.1$.
However, using $\chi_{p}$ alone, 
\textit{even after performing the parameter recovery} there are no simple criteria to determine when 
precession is observed. A natural choice might require that the 90\% confidence interval for $\chi_p$ exclude zero, 
but this will \textit{always} be the case, primarily due to the shape of the prior. Furthermore, even though we know all of the parameters \emph{a priori}, it is impossible to determine whether precession will be observable without 
generating the waveform and performing the parameter recovery.  

The precession SNR solves these problems. A value of $\rho_p > 2.1 $ tells us immediately that precession is observable. 
Most significantly, the expected precession SNR $\rho_{p}$ (red lines on the middle plot) can be calculated 
directly from the signal parameters; no detailed parameter estimation analysis is necessary.  Thus, for the first
time, we are able to identify immediately whether precession would be measurable in a given configuration. 
We see in Fig. ~\ref{fig:chi_p_and_rho_p} that for each inclination, the true value for  $\rho_p $ lies within the recovered 
90\% credible interval however the posterior is not centred around the true value. This is due to selection and prior effects.
In Ref.~\cite{Green:2019_prec_pe}, we investigate these selection effects as well as providing a detailed exploration
of the observability of precession over the  parameter space of masses, spins and binary orientation.

Fig.~\ref{fig:chi_p_and_rho_p} also shows the distribution of $\rho_{p}$ for the BBH merger signals that
were observed in the aLIGO and AdV O1 and O2 runs \cite{LIGOScientific:2018mvr}.  No evidence of precession was found in
these signals \cite{LIGOScientific:2018jsj}, as is made clear from the recovery of $\rho_p$.  There are several cases
where the distribution extends to higher values, but the median never exceeds the 2.1 threshold.
These results demonstrate the efficacy of $\rho_p$. 

{ \ }
\textbf{When will we observe precession?}
We can use the observation of precession to distinguish different binary formation scenarios. 
The precession SNR makes it straightforward to perform an in-depth investigation of various models 
and identify the fraction of signals for which precession effects will be observable. Such a study was not previously possible,
due to the difficulty in classifying observability of precession. Instead, limited investigations of the parameter space have
been performed~\cite{Vitale:2014mka}, or inferences of the distributions for the spin magnitudes and orientations obtained 
\cite{Wysocki:2018mpo, Talbot:2017yur}, again with a limited sample size.  

We investigate nine astrophysically-motivated populations of black hole binaries, comprised of three distributions of spin 
magnitude, and three distributions of spin orientation for the individual black holes in the binary.  
The spin-magnitude distributions are those used in Refs.~\cite{Farr:2017uvj, Farr:2017gtv, Tiwari:2018qch}: 
\emph{low} and \emph{high} are triangular, peaked either at zero or extremal spin, and \emph{flat} is a uniform distribution 
between zero and one. 
The spin-orientation is characterized by the distribution for the angle $\sigma$ between the black hole's 
spin and the orbital angular momentum: \emph{aligned} is a triangular distribution in $ \cos{\sigma} $, 
which peaks at 1 and can take values $0.85<\cos{\sigma}<1.0$, ($\sigma \lesssim30\degree$); 
\emph{precessing} is triangular in $\cos{\sigma}$ peaked at 0, with values $-0.15<\cos{\sigma}<0.15$, 
($80\degree \lesssim \sigma \lesssim 100\degree$); 
\emph{isotropic} is uniform in $\cos{\sigma}$ between $-1$ and $1$.  
For each population, we generate $10^5$ binaries with masses drawn from a power law mass distribution 
with $p(m_{1}) \propto m^{-2.35}$,
and $p(m_2)$ uniform in $m_2$ between $5\,M_\odot$ and $m_1$ (as in \cite{Tiwari:2018qch}),
and distributed uniformly in volume and binary orientation. 

\begin{table}
	\begin{center}
		\begin{tabular}{c || c | c || c | c || c | c ||}
			& \multicolumn{2}{c || }{Aligned} & \multicolumn{2}{c || }{Isotropic} & \multicolumn{2}{c || }{Precessing} \\
			\hline
			Low & 0.043 & \cellcolor{gray!40} 0.644 & 0.151 & \cellcolor{gray!40} 0.194 & 0.173 & \cellcolor{gray!40}  0.150 \\
			\hline
			Flat & 0.077 & \cellcolor{gray!40} 0.448 & 0.276 & \cellcolor{gray!40} 0.040 & 0.327 & \cellcolor{gray!40} 0.019 \\
			\hline
			High & 0.105 & \cellcolor{gray!40} 0.331 & 0.354 & \cellcolor{gray!40} 0.013 & 0.412 & \cellcolor{gray!40} 0.005 \\
			\hline
		\end{tabular}
	\end{center}
	\caption{The probability of observing precession, $\rho_{p} > 2.1$, for an observed binary (white) from each spin 
	distribution and the probability of \emph{not} observing precession in 10 random draws (grey)
	from each spin distribution.
	\label{tab:precession_prob}
	}
\end{table}

Table \ref{tab:precession_prob} shows the probability of observing precession, $\rho_{p} > 2.1$,
in a single event drawn from each of the nine populations, observed with O2 senstivity.  
As expected, we are most likely to observe precession when the black holes have high spins that lie preferentially in the orbital
plane (\emph{high-precessing} configurations) and least likely for black holes with low spins, or with
spins preferentially aligned with the orbital angular momentum (\emph{low-aligned} configurations).
Given that precession has not been observed in GW detections to date, we are able to restrict the
spin distribution. Table \ref{tab:precession_prob} shows the probability of detecting ten signals
with no observable precession from each of the nine spin distributions.  Based on precession measurements alone, we
strongly disfavour all \emph{precessing} distributions.  Although these are already considered astrophysically unlikely, there
are models that predict preferentially in-plane spins \cite{Antonini:2017tgo, Rodriguez:2018jqu}.  We also
disfavour \emph{isotropic} spins with \emph{flat} or \emph{high} magnitudes.  Thus, the lack of observed precession
points towards low spins, or spins preferentially aligned with the orbital angular momentum.

\begin{figure*}[t!]
	\centering
	\includegraphics[scale=0.3]{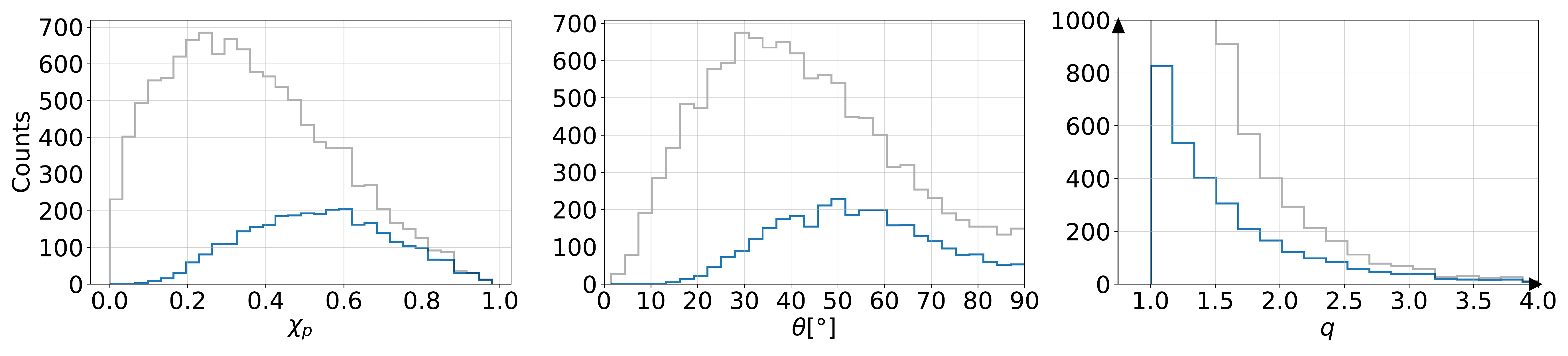}
	\caption{The distribution of $\chi_{p}$, $\theta$ and $q$
	for  observable binaries (grey), and those with measurable precession (blue), assuming
	a low isotropic spin distribution.  $\theta$ is the inclination angle folded to $[0, \pi/2]$.
	The y-axis labels the number of observed events in each bin, out of 1$0^{5}$ simulated signals with low isotropic spins.
	}
	\label{fig:precession_population}
\end{figure*}

Previous constraints on spins have primarily been provided by considering the measurable aligned-spin
component~\cite{Farr:2017uvj, Farr:2017gtv, Tiwari:2018qch, LIGOScientific:2018mvr} 
and provide strong evidence against all but \emph{low} \emph{aligned} or \emph{iostropic} distributions, with
\emph{low isotropic} spins preferred.  Combining the aligned spin and precession results will further restrict the spin 
distribution consistent with GW observations, and will likely require spin magnitudes even smaller than our 
\textit{low} distribution (see also Ref.~\cite{LIGOScientific:2018jsj}).

{ \ }
\textbf{Where will we observe precession?}
We are able to identify, for the first time, the regions of parameter space that lead to signals with observable precession.  In Fig.~\ref{fig:precession_population},
we show the expected distribution of the precessing spin $\chi_{p}$, binary orientation $\theta$ and mass ratio $q$ for observable binaries and binaries with observable precession, $\rho_p > 2.1$, assuming a \textit{low isotropic} distribution of spins.
We identify clear regions of the parameter space where precession is more likely to be observed: large
values of $\chi_{p}$, binaries that are close to edge-on, $\theta > 45^{\circ}$, and systems with high mass ratio.
Regions where the chance of observing precession is close to zero include binaries with $\chi_{p} < 0.2$ or
where the total angular momentum is within $20\degree$ of the line of sight.  These results are consistent with expectations
based upon smaller studies using detailed parameter estimation techniques \cite{Vitale:2014mka}.  
We also note that most observations of precession will be in comparable-mass binaries, i.e., 
$q\leq 2$. This is a surprising, new result.  It is well known that precession is more easily measured at higher mass ratios 
\cite{Apostolatos:1994mx}, which is confirmed by our study: precession is observed in $<$12\% of detections with $q<2$, but $>$35\% for $q>2$. However, with $\sim90\%$ of observations expected to have
$q<2$,
these vastly outnumber the higher-mass-ratio observations, and we find that $\sim$75\% of precession observations will 
come from detections of binaries with $q<2$. 

\textbf{Discussion:}
In this letter we have used a simple method to identify when precession is measurable in a compact binary
GW signal. The gravitational waveform is well approximated by the first two harmonics in a power series expansion in
the tangent of the half-opening angle \cite{Fairhurst:2019_2harm}, and the unambiguous observation of precession requires the
identification of both of these harmonics in the data.  The precession SNR $\rho_p$ is a simple measure of this observability.
We have demonstrated the efficacy of $\rho_{p}$ through parameter estimation studies and also
provided the distributions of $\rho_p$ for the aLIGO-AdV observations to date.
Using our definition of precession SNR, we have identified how often precession will be observed for a variety of
potential astrophysical spin distributions. For the most likely distribution, based on current observations (\emph{low-aligned})
there is a 83\% chance that precession will be measured after $\sim$40 observations, and is therefore likely to be observed
during the current third aLIGO-AdV observing run (O3). The non-measurement of precession by the end of O3 would
place much stronger constraints on spin orientations and magnitudes.

The precession SNR has many applications. Most immediately, it allows us to determine the measurability of
precession in a system \emph{without performing computationally expensive parameter estimation}.
This allows us to, e.g., easily fold precession
information into population analyses of black-hole binaries.  In future work, we will explore whether the value of $\rho_{p}$
can be used to predict the measured $\chi_{p}$ distribution.  The precessing SNR also gives us a simple way to
identify regions of the parameter space where precession is important, a necessary first step in extending
existing GW searches to explicitly use precessing waveforms \cite{Harry:2016ijz}.

\textbf{Note:} Since the first submission of this Letter, $\rho_p$ has been used in the analysis of the aLIGO-AdV 
signals GW190412~\cite{LIGOScientific:2020stg} and GW190814~\citep{Abbott:2020khf}.

\textbf{Acknowledgements}
We thank Eleanor Hamilton, Ian Harry, Alistair Muir, Francesco Pannarale, Vivien Raymond, Cory
Thomas, and Vaibhav Tiwari for discussions.
This work was supported by Science and Technology Facilities Council (STFC) grant ST/L000962/1 and European
Research Council Consolidator Grant 647839.
Computational resources were provided by Cardiff University, funded by
STFC support for UK involvement in aLIGO operations.

\bibliographystyle{unsrt_LIGO}
\bibliography{main}

\end{document}